\documentclass[a4paper]{article}

\usepackage{amsmath,amssymb,amsthm}

\newcommand{\vtri}{\boldsymbol{\vartriangle}}

\begin{document}

\title{\bf  
 A Possible Mechanism of Biological Memories 
\break 
 in terms of Quantum Fluids
}

\author{Tsunehiro Kobayashi\footnote{E-mail: 
kobayash@a.tsukuba-tech.ac.jp} \\
{\footnotesize\it Department of General Education 
for the Hearing Impaired,}
{\footnotesize\it Tsukuba College of Technology}\\
{\footnotesize\it Ibaraki 305-0005, Japan}}

\date{}

\maketitle

\begin{abstract}

A mechanism of memories, especially biological memories, is studied 
in terms of quantum fluids. 
Two-dimensional flows in central potentials 
$V_a(\rho)=-a^2g_a\rho^{2(a-1)}$ 
($a\not=0$ and $\rho=\sqrt{x^2+y^2}$) 
have zero-energy eigenstates 
that degenerate infinitely for all $a$. 
It is shown that 
stable standing waves constructed from the zero-energy flows are 
confined in various types of polygons which can be the minimum units 
of memory systems. 
Vortex patterns awoken in the units by stimuli correspond to 
the memories of the stimuli. 
This memory system is not a system for preserving 
memories as usual but that for awaking memories.  
The system has interesting properties; 
(i) the absolute economy as for the energy consumption, 
(ii) the infinite variety 
for a huge number of memories, 
 (iii) the perfect recovery of the system from any disturbances by stimuli, 
 and (iv) the large flexibility in the construction of the system. 
A process for thinking is also proposed in terms of this memory system.

\end{abstract}

\thispagestyle{empty}

\setcounter{page}{0}

\pagebreak
\hfil\break
{\bf 1. Introduction}

Researches for the mechanism of brains are extensively going on from 
many different 
experimental and theoretical standpoints, 
and many interesting results have been reported. 
At present, however, these approaches are still not enough to understand 
the mechanism of brains from the viewpoint of a fundamental theory of physics. 
The first step of the understanding seems to be the elucidation of 
the mechanism of memories in brains as a physical process.  
The insides of our bodies are mainly occupied by many kinds of fluids, 
and the total unification of  
many organs are also preserved mainly by such fluids. 
It is not a peculiar idea that the mechanism in 
brains is governed by some kind of flows of such fluids. 
In this sense it is quite natural that 
the dynamics of flows (hydrodynamics) is one of 
the fundamental dynamics in our bodies. 
On the other hand 
the dynamics in brains does not seem to be a completely 
deterministic dynamics like classical dynamics, 
but some kind of probability seems to play 
an important role in the mechanism of thinking. 
Considering the coexistence of the probability and 
the fundamental principle of superposition in wave dynamics, 
we may invoke some kind of quantum process related to 
hydrodynamics to describe the dynamics in brains.  
Such a hydrodynamical approach of quantum mechanics was vigorously examined 
in the early stage of the
development of quantum mechanics [1-8], 
and then some fundamental properties 
of the quantum flows 
like vortices were investigated [9-13]. 
Recently interesting solutions that have exactly zero-energy eigenvalue 
and degenerate infinitely 
have been found in two-dimensional quantum flows [14-16]. 
From the viewpoint of the energy consumption in the construction 
and the preservation of the 
mechanism of brains such zero-energy flows can be very useful tools. 
Furthermore the infinite degeneracy 
can provide the infinite variety of vortex patterns [15-17] 
corresponding to a huge number of memories. 
Note that the vortices 
have recently been observed 
in many phenomena such as condensed matters [18-20], 
non-neutral plasma [21-24] 
and Bose-Einstein gases [25-29]. 
Here we show a possibility of the mechanism of memories 
in terms of the zero-energy flows. 
It is shown that the memory system proposed here is not a system 
for preserving memories as usual but that for awaking memories, 
and it has quite suitable properties to represent biological memory 
systems. 
\vskip10pt
\hfil\break
{\bf 2. A short review on zero-energy flows in two-dimensional 
Schr$\ddot {\rm o}$dinger equations} 

In this section we briefly study two-dimensional (2D) quantum flows 
and vortices. 
It has been shown that 
2D Schr$\ddot {\rm o}$dinger equations with central potentials 
$V_a(\rho)=-a^2g_a\rho^{2(a-1)}$ ($a\not=0$ and  
$\rho=\sqrt{x^2+y^2}$), of which 
eigenvalue problems with the energy eigenvalue ${\cal E}$ 
are written as
\begin{equation}
 [-{\hbar^2 \over 2m}\vtri +V_a(\rho)]\ \psi(x,y) 
  = {\cal E}\ \psi(x,y)
  \label{0}
\end{equation}
where 
$
{\boldsymbol{\triangle}}=\partial^2/ \partial x^2+\partial^2 / \partial y^2,
$  
have zero-energy (${\cal E}=0$) eigenstates [14-16]. 
Note here that in this equation the mass $m$ and the coordinate vector $(x,y)$ 
can represent not only those of the single particle but those of the centre of 
mass for a many particle system as well~\cite{k2}. 
It has also been shown that, 
as far as the zero-energy eigenstates ($\psi_0$) are concerned, 
the Schr$\ddot {\rm o}$dinger equations for all $a$ can be reduced to 
the following equation 
in terms of the conformal transformations 
$\zeta_a=z^a$ with $z=x+iy$; 
\begin{equation}
[-{\hbar^2 \over 2m}{\boldsymbol{\triangle}}_a-g_a]\ \ \psi_0 (u_a,v_a)=0, 
 \label{1}
\end{equation}
where
$ 
{\boldsymbol{\triangle}}_a=\partial^2/ \partial u_a^2+\partial^2/ \partial v_a^2,
$  
using the variables defined by 
the relation $\zeta_a=u_a+iv_a$~\cite{k1,ks1}. 
That is to say, the zero-energy eigenstates for all the different numbers of $a$ 
are described by the same plane-wave 
solutions in the $\zeta_a$ space. 
Furthermore it is easily shown 
that the zero-energy states degenerate infinitely.
Let us consider the case for $a>0$ and $g_a>0$. 
Putting the function $f^\pm_n (u_a;v_a)e^{\pm ik_a u_a}$ with 
$k_a=\sqrt{2mg_a}/\hbar $ 
into (2), 
where $f^\pm_n (u_a;v_a)$ are polynomials of degree $n$ ($n=0,1,2,\cdots $), 
we obtain the equation for the polynomials 
\begin{equation}
[{\boldsymbol{\triangle}}_a \pm2ik_a{\partial \over \partial u_a}]f^\pm_n (u_a;v_a)=0.
\label{2}
\end{equation} 
Note that from the above equations we can easily see the relation  
$(f^-_n (u_a;v_a))^*=f^+_n (u_a;v_a)$ for all $a$ and $n$. 
General forms of the polynomials have been obtained by using the solutions 
in the $a=2$ case 
(2D parabolic potential barrier (2D PPB))~\cite{k1,ks1}. 
Since all the solutions have the factors $e^{\pm ik_a u_a}$ 
or $e^{\pm ik_a v_a}$, we see that 
the zero-energy states describe stationary flows~\cite{sk4,k1,ks1}. 
Taking account of the direction of incoming flows in the 
$\zeta_a$ plane that is expressed by the angle $\alpha$ 
to the $u_a$ axis, the general eigenfunctions 
with zero-energy are written as arbitrary linear combinations of 
eigenfunctions included in two infinite 
series of $\left\{ \psi_{0n}^{\pm (u)}(u_a(\alpha);v_a(\alpha))  \right\}$ 
for $n=0,1,2,\cdots$, where 
\begin{equation} 
\psi_{0n}^{\pm (u)}(u_a(\alpha);v_a(\alpha))=
f_n^\pm(u_a(\alpha);v_a(\alpha))e^{\pm ik_a u_a(\alpha)}, 
\label{A}
\end{equation}
$u_a(\alpha) =u_a {\rm cos}\alpha +v_a {\rm sin}\alpha $, 
and $0<\alpha \leq 2\pi$. 
(For the details, see the sections II and III of Ref. 15.)
We briefly comment on states having non-zero energy eigenvalues. 
Energy eigenvalues are in general complex values, 
and the states with complex energies can be understood 
to be resonance states which are not stable and decay 
in some time scale~\cite{Bohm}. 
Exact solutions not only for zero-energy but for non-zero energies as well 
have been solved in 2D PPB~\cite{sk4,sk}.    
Note here that we consider  
the motions of the $z$ direction perpendicular to the $xy$ plane 
to be free motions represented by $e^{\pm ik_z z}$. 
We should, therefore, notice that 
the total energies $E_T$ of the flows are given by 
$E_T=E_z$, where $E_z$ are the energies of the free motions in the $z$ direction. 
We shall, however, not take account of the motions in the $z$ direction 
in the following discussions 
on the motions in the $\zeta_a$ plane.

Note that the zero-energy eigenfunctions can not be normalized as same as those 
in scattering processes~\cite{Bohm}. 
Actually it has been shown that all the states 
for zero and non-zero energies are eigenstates 
in the conjugate spaces of the Gel'fand triplets [13-16]. 
This fact means that the probability density ($\rho(t,x,y)=|\psi(t,x,y)|^2$) 
and the probability current 
(${\boldsymbol{j}}(t,x,y)={\rm Re}[\psi(t,x,y)^*(-i\hbar \nabla)\psi(t,x,y)]$) 
lose the meanings. 
We, however, see that 
the velocity defined by  
$ 
 {\boldsymbol{v}}={\boldsymbol{j}}(t,x,y)/\rho(t,x,y), 
$ 
can have the well-defined meaning, 
because the ambiguity arising from the normalization 
is cancelled in the definition. 
The velocity was widely discussed in early stage of quantum mechanics
 [1-8]. 
It is also well-known that 
vortices appear at the zero points of 
the density, that is, the nodal points of the wave functions
 [9-13]. 
Let us here remember the fact that the zero-energy states have
the infinite degeneracy, and therefore 
we can construct wave functions having 
the nodal points at almost arbitrary positions in terms of 
linear combinations of the zero-energy states. 
Actually various vortex patterns have been studied [15-17]. 
It should be emphasized that 
we can study the vortex problems 
for all $V_a(\rho)$ as a problem with the constant potential $V_a=-g_a$ 
in the $\zeta_a$ plane, 
because the conformal transformations do not change 
fundamental properties of vortices such as 
the numbers of vortices 
and 
the strengths of vortices $\Gamma$. 
In quantum mechanical phenomena it is shown that 
the circulation $\Gamma=\oint_C {\boldsymbol{v}} \cdot d{\bf s}$ 
is quantized such that 
$\Gamma=2\pi l\hbar/m$, where $l$ is an integer
 [9-14]. 
Vortex patterns in the $(x,y)$ plane can be obtained by the inverse 
transformations of the conformal mappings~\cite{k1,ks1}. 
\vskip10pt
\hfil\break
{\bf 3. Standing waves and vortex patterns} 

Let us start with the study of zero-energy standing waves that will play 
an important role in the following discussions. 
Considering the relation $(f^-_n (u_a;v_a))^*=f^+_n (u_a;v_a)$, 
we can easily make 
two infinite series of standing waves 
in terms of the zero-energy series 
$\left\{ \psi_{0n}^{\pm (u)}  \right\}$ 
as follows;
\begin{align}
\psi_{0n}^{+{\rm S}(u)}&={1 \over 2}(\psi_{0n}^{+ (u)} + \psi_{0n}^{- (u)}) 
 =f_n^{{\rm Re}(u)} {\rm cos}k_a u_a(\alpha )-f_n^{{\rm Im}(u)} {\rm sin}k_a u_a(\alpha )
, \nonumber \\
\psi_{0n}^{-{\rm S}(u)}&={1 \over 2}(\psi_{0n}^{+ (u)} - \psi_{0n}^{- (u)}) 
 =i[f_n^{{\rm Re}(u)} {\rm sin}k_a u_a(\alpha )+f_n^{{\rm Im}(u)} {\rm cos}k_a u_a(\alpha )],
\label{T}
\end{align} 
where $f_n^{{\rm Re}(u)}$ and $f_n^{{\rm Im}(u)}$, 
respectively, stand for 
the real and imaginary parts of $f_n^{+}$. 
In order to simplify our argument 
we pick up only four standing waves that are constructed in terms of 
the linear combinations of 
the flows with the lowest order, that is, $e^{\pm i(k_u u_a\pm k_v v_a)}$ 
corresponding to the choices of $\alpha =\theta_a, \ \pi \pm \theta_a$ 
and $2\pi-\theta_a$, where 
$k_u=k_a{\rm cos}\theta_a$ and $k_v=k_a{\rm sin}\theta_a$ 
with $0<\theta_a\leq 2\pi$, such that 
\begin{equation} 
\psi_{cc}={\rm cos}k_u u_a\cdot {\rm cos}k_v v_a,\ \ 
\psi_{ss}={\rm sin}k_u u_a\cdot {\rm sin}k_v v_a,\ \ 
\psi_{sc}=i{\rm sin}k_u u_a\cdot {\rm cos}k_v v_a,\ \ 
\psi_{cs}=i{\rm cos}k_u u_a\cdot {\rm sin}k_v v_a.
\label{3}
\end{equation} 
These wave functions can be confined in quadrangles of the $\zeta_a$ plane, 
which are surrounded by rigid fences described by infinite potentials, 
because we can take zero as boundary values 
on the sides of the quadrangles. 
For example 
$\psi_{cc}={\rm cos}k_u u_a\ {\rm cos}k_v u_a$ becomes zero on the lines 
where either $k_u u_a=(2m+1)\pi/2$ or $k_v v_a=(2n+1)\pi/2$ 
($m$ and $n$ are integers) is satisfied. 
In the cases where $a$ are positive integers the quadrangles 
in the $\zeta_a$ plane are represented by $4a$-gons 
in the original $xy$ plane. 
Now we get the standing wave states confined in the special regions 
which are described by the quadrangles in the $\zeta_a$ plane. 
It should be noticed that the quadrangles have a wide variety 
of the sizes which are determined 
by the numbers of wave lengths contained in the quadrangles. 
That is to say, the number $N_u$ for the $u$ direction and 
that $N_v$ for the $v$ direction are 
free parameters, where $N_u$ and $N_v$ are positive integers or 
positive half-integers.

Now let us discuss vortex patterns made from 
the zero-energy flows. 
It is known that vortex patterns are one of topological properties 
of flows, and then they can be stable in many physical phenomena. 
Following the above consideration, 
we can prepare many different standing waves depending on  
the sizes of quadrangles. 
The standing waves, of course, have no vortex, because ${\boldsymbol{j}}=0$. 
Let us study an example of a quadrangle where a standing wave 
$\psi_{cc}$ with $N_u=1/2$ and $N_v=3/2$ 
is confined. 
We would like to investigate what happens 
when a flow is put into the quadrangle.  
Let us put a different standing wave, for example, 
the standing wave with the wave number vector $\vec q=(q_u,q_v)$, 
which is described by the wave function 
$\psi_{sc}=i{\rm sin}q_u u_a\cdot {\rm cos}q_v v_a$, 
into the quadrangle prepared. 
Note that in the process where the new flow is added we can use 
the free motions of the $z$ direction. 
We see that the ranges of $u$ and $v$ in the quadrangle are given by 
$|u| < u_B\equiv \pi/2k_u$ and $ |v| <v_B\equiv 3\pi/2k_v$. 
(Hereafter we omit the suffix $a$ from $u_a$, $v_a$, $k_a$ 
and etc.) 
The total wave function is written as 
\begin{equation}
\Psi={\rm cos}k_u u\cdot {\rm cos}k_v v +iC{\rm sin}q_u u\cdot {\rm cos}q_v v, 
\label{5}
\end{equation} 
where $C$ is taken to be a real number 
for the simplicity of the present discussion.  
In order that the inserted wave $\psi_{sc}$ is stable in the quadrangle,    
the  wave function of the inserted wave must vanish 
on the boundaries of the quadrangle, that is, on the lines fulfilling 
$|u|=u_B$ 
or $|v|=v_B$. 
This constraint gives us the relations 
\begin{equation}
q_u=2(L_u+1) k_u\ \ {\rm and} \ \ q_v=(2L_v+1)k_v/3,
\label{6}
\end{equation} 
where $L_u$ and $L_v$ are zero or positive integers. 
Let us investigate vortices in the quadrangle. 
From the fact that $\psi_{cc}$  
has two nodal lines fulfilling $v=\pm v_B/3$ in the $\zeta$ plane 
and on the other hand $\psi_{sc}$ has 
nodal lines fulfilling $u=n\pi/q_u$ or $v=(2m+1)\pi/2q_v$ 
($n$ and $m$ are integers),  
we see that vortices  
appear at the nodal points of the total wave function $(u_V,v_V)$, where  
\begin{equation}
u_V=n\pi/q_u=\pm nu_B/(L_u+1), \ \ \ {\rm and}\ \ \ 
v_V=\pm v_B/3.  
\label{7}
\end{equation} 
Considering the constraint $|u|<u_B$, we obtain the following results; 
    
in the case of $L_u=0$ two vortices appear at $(0,\pm v_B/3)$ for $n=0$, 
    
in the case of $L_u=1$ six vortices appear at $(0,\pm v_B/3)$ for $n=0$ 
and $(\pm u_B/2,\pm v_B/3)$ for $n=1$, 
\hfil\break
and generally 
    
$2(L_u+2)$ vortices appear on the lines of $v=\pm v_B/3$ for $L_u>0$. 
\hfil\break
In general cases the following results are found out: 
\hfil\break
(I){\it In this vortex-formation process we find out the selection rule 
for the inserted wave numbers $\vec q$ such that only the waves with 
$q_u=(L_u+1) k_u/N_u$ and $q_v=(2L_v+1) k_v/2N_v$ can format 
some stable vortex-patterns in the above example.} 
In general the ratios of $q_u/k_u$ and $q_v/k_v$ must be rational 
numbers. 
\hfil\break
(II) {\it From the observation of the number of the vortices 
we can read off the number $L_u$ (or $L_v$).} 
\hfil\break 
Furthermore we can add one more important fact: 
\hfil\break
(III) {\it In the processes putting standing waves into the quadrangle 
the number $C$ that represents the magnitude of the inserted wave 
play no role in the determination of the vortex patterns.}  
\vskip10pt
\hfil\break
{\bf 4. Mechanism of memories} 

Here let us consider the mechanism of memories in terms of 
the vortex-formation process. 
In the present framework 
we consider that every vortex pattern corresponds to a memory. 
In this viewpoint 
a quadrangle confining a standing wave is not a memory corresponding 
to a special item,  
but  
the system in which a vortex pattern is awoken by a flow 
induced by a stimulus. 
A quadrangle can awake different vortex patterns corresponding to 
different wave numbers expressed by the relations 
like $q_u=(L_u+1) k_u/N_u$  in the example of the section 3. 
We may visualize the following process; 
a stimulus corresponding to an item is taken by some organs 
and converted to flows. 
The flows reach at boundaries of many quadrangles 
with different sizes ($4a$-gons in the real space).  
Every quadrangle analyses the flow in terms of the standing waves 
given by (6). 
Notice that 
this analysis for the inserted flow is nothing but  
Fourier analysis in the $\zeta_a$ plane, of which boundaries 
are given by the lines $u=\pm u_B$ for 
$u$ and $v=\pm v_B$ for $v$. 
If the flow does not contain the waves with the wave numbers 
fitting to the selection rule (I), 
the flow cannot  make 
any stable vortex pattern. 
Only the flows containing the wave numbers fitting to the rule (I) make 
stable vortex patterns in the quadrangles. 
Thus one stimulus sent to many quadrangles 
awakes various vortex patterns in those quadrangles. 
We may consider that the set of these vortex patterns made in 
the different quadrangles 
represents the memory corresponding to the stimulus. 
As noted in (III), the magnitude of the stimulus does not affect 
the vortex formation at all. 

Let us see the characteristic features of this memory system. 
From the mechanical standpoint the fundamental properties 
of a quadrangle 
are determined by the parameters of the potential inside the quadrangle, 
that is, $a$ and $g_a$, 
where $a$ determines the form ($4a$-gon) 
of the $xy$ plane and $g_a$ does the fundamental wave number $k_a$. 
It must be noticed that further steps are needed 
for making a quadrangle confining a standing wave. 
One is 
the division of the fundamental wave number 
between two directions $u_a$ and $v_a$, 
following the relation $k_a^2=k_u^2+k_v^2$, that is, $\vec k_a=(k_u,k_v)$ is  
the wave number vector in the $\zeta_a$ plane.  
In order to determine the actual lengths 
of the sides of the quadrangle, we need one more step 
for the decision of two numbers $N_u$ and $N_v$ which determine 
the lengths of the sides of the quadrangle. 
Summarizing the above argument, 
the following three steps are needed to set up a quadrangle; 

(1) the choice of $a$ and $g_a$ (the determination of the potential 
inside the quadrangle), 

(2) the division of $k_a$ 
between $k_u$ and $k_v$ 
(the determination of the wave vector $\vec k_a$ 
confined 

in the quadrangle), and 

(3) the choice of the numbers of the wave lengths 
contained in the quadrangle, $N_u$ and $N_v$ 

(the determination of the lengths of the sides of the quadrangle). 
\hfil\break
Through these processes a quadrangle confining  a definite standing wave 
is determined. 
We see that in these processes we have five parameters, that is, $a $, $g_a$, 
$N_u$, $N_v$ and one more parameter $\theta_a$ to determine $\vec k_a$ 
such that $k_u=k_a{\rm cos}\theta_a$ and $k_v=k_a{\rm sin}\theta_a$. 
The angle parameter $0<\theta_a\leq 2\pi$ is a continuous 
free-parameter, and therefore the two wave numbers $k_u$ and $k_v$ 
can be also continuous numbers limited by $k_a$. 
This fact means that 
from one fundamental wave number $k_a$ 
we can make quadrangles having arbitrary values as for the ration of 
the lengths of two perpendicular sides, which is given by tan$\theta_a$.   
Taking also account of the free parameters $N_u$ and $N_v$, we have large freedom 
to the determination of the sizes and forms of quadrangles. 
We can, thus, provide quadrangles which can correspond to almost 
arbitrary inserted wave number vectors $\vec q$ 
even if $\vec q$ must be satisfied 
by the selection rule given in (I) to make stable vortex patterns 
in the quadrangles.

Let us continue to discuss about the mechanism for making memories. 
As already noted, the quadrangles are not the mechanism for 
preserving memories but that for awaking memories. 
One quadrangle characterized by a set of the four parameters 
$k_a,\ \theta_a$, $N_u$ and $N_v$ 
can fundamentally produce infinite numbers of 
different vortex patterns corresponding to the differences of 
the wave numbers of the inserted waves expressed by the relations 
$q_u=L_uk_u/N_u$ and $q_v=L_vk_v/N_v$, 
where $L_u$ and $L_v$ are positive integers or positive half-integers. 
It should once more be emphasized that one quadrangle 
awakes only one pattern for a stimulus but the pattern 
corresponding to the stimulus is generally different 
from others. 
Considering the limit of the discrimination of the vortex patterns, 
each quadrangle may have a certain maximum values $(L_{\rm max})$ 
for $L_u$ and $L_v$. 
In such a case, for the system where $N_q$ number of quadrangles cooperate 
we can roughly estimate the  
number ($N_{VP}$) of the different vortex patterns 
that can be distingushed by the system as  
$ 
N_{VP}=\prod_{i=1}^{N_q} ( L_{u,{\rm max}}\times L_{v,{\rm max}})_i. 
$ 
Provided that the mean number of 
$( L_{u,{\rm max}}\times L_{v,{\rm max}})_i$ is 10 and 
$N_q$ is a large number, 
we obtain a huge number of the distinguishable vortex patterns 
counted to be $N_{VP}\sim 10^{N_q}$. 
Even if the mean number is 2 and $N_q$ is not a very large number, e.g. 
100, we still have a huge number $2^{100}\sim 10^{30}$. 
From this fact we may consider that a total system consisting of a huge number 
of quadrangles is divided into many small blocks, 
and each block plays a fairly independent role 
in the total system. 

We would like here briefly to note on the role of 
the higher order standing waves given in (5). 
Taking account of the freedom introduced in linear combinations 
of those standing waves, we will find out that 
the boundary conditions are expressed by some complicated relations 
for $u_a$ and $v_a$.  
On the other hand, from (5) we see that 
all the terms of the standing waves have 
one of the four factors  ${\rm cos}k_uu_a$, 
${\rm sin}k_uu_a$, ${\rm cos}k_vv_a$ and ${\rm sin}k_vv_a$. 
Therefore, as far as the discussions of vortex patterns are concerned, 
we can perform arguments similar to those done in the lowest cases. 
In actual situations, however, the fences of quadrangles ($4a$-gons in real spaces) 
will be made of some 
membranes in living beings. 
They are, of course, not perfectly rigid. 
In such cases the wave functions do not generally vanish at the boundaries 
of the quadrangles. 
Boundaries are, therefore, not given by strict quadrangles but 
by some deformed forms, and the values of wave functions on the boundaries  
are not 
strictly 0 but some values determined by the continuity conditions of wave functions. 
The infinite degeneracy of the standing waves can possibly play a role to solve 
the problem. 
That is to say, we will be able to provide various forms corresponding 
to various boundary conditions in terms of the infinite degeneracy. 
The vortex positions must also be changed. 
By gathering the terms with the same factor (cos or sin) 
discussions 
similar to those done in the lowest order cases will be also performed here. 
This infinite degeneracy and the potential parameters $a$ and $g_a$ 
will bring a large flexibility in the construction of 
the minimum units of the memory system. 
\vskip10pt
\hfil\break
{\bf 5. Interesting properties of the system composed of the zero-energy flows} 

We have used only the zero-energy states for making the memory system. 
We would like to point out 
four interesting properties of the present system composed of 
the zero-energy flows~\cite{k2}; 
\hfil\break
(i) the absolute economy as for the energy consumption in the construction 
of the memory system and also the preservation of the system, 
where no flow in the $z$ direction should be taken, and therefore 
the total energy $E_T=0$. 
\hfil\break
(ii) the almost infinite variety of the vortex patterns provided for 
a huge number of memories, and 
\hfil\break
(iii) the perfect recovery of the memory system from shocks induced by 
stimuli, 
because the zero-energy ($E_T=0$) 
states have no time dependence at all, and therefore 
the standing waves in the minimum units of the system 
are perfectly recovered even if any disturbances induced 
by flows decaying or diffusing in some time scale 
are taken place~\cite{k2}. 
\hfil\break
These properties are suitable 
for interpreting the mechanism of memories in brains. 
That is to say, stimuli taken in sense organs are analyzed 
in terms of the states 
including both of zero- and nonzero-energy states 
in the conjugate spaces of Gel'fand triplets, 
and after some time scale 
only the zero-energy flows selected by the rule (I) 
remain in the minimum units and make stable vortex patterns. 
The vortex patterns are sent to other places by the free motions 
of the inserted waves in the direction perpendicular to the $xy$ plane, 
and then every unit perfectly recover the initial state. 
We may point out one more important property; 
\hfil\break
(iv) the large flexibility due to 
the infinite degeneracy and the parameters $a$ and $g_a$ in the potentials 
is provided in making the minimum units like the quadrangles. 
\vskip10pt
\hfil\break
{\bf 6.  A comment on thinking processes} 

Finally we would like to present a conjecture about thinking processes. 
At present we do not know how the mechanism of memories presented here  
does work in thinking processes (algorithm of thinking processes). 
We shall here point out a possibility that the memory system 
proposed here plays a role to distinguish accustomed 
stimuli, which have been repeatedly experienced in the past, 
from many stimuli included in an item caught by sense organs. 
We may consider that the same stimulus repeatedly given in our experience 
provokes organizations in brains to make 
a memory system consisting of 
many minimum units like quadrangles 
(possibly neurons in our brains) for selecting 
the very stimulus among many stimuli.  
Some systems like autonomic nervous system are made in the womb of mother. 
Such a memory system provided for the selection of some special stimuli will be made  
in a rather small block of the total system. 

Now we can consider the following process for thinking: 
When an item is caught by some sense organs, the stimuli taken by the organs 
will be gathered in a special place of brains, 
where the stimuli are converted into signals that fit 
to the detection by the memory system. 
From the place the signals are delivered 
to many blocks, each of which consist of a certain number of 
minimum units. 
The signal sent to one of the blocks are analyzed 
by each unit in the block in terms of 
the standing waves, and then 
every unit selects the standing waves, following the selection rule (I).
If there exist wave numbers fitting to the rule, the unit make a vortex pattern.  
Thus the signal sent to the block 
is converted to a set of vortex patterns. 
This set represents a recognition of an accustomed experience. 
Thus the stimuli caught by the organs are represented 
by a set of such recognitions . 
The recognitions characterized 
by various vortex patterns are once more delivered 
to a certain number of new blocks 
(call them higher blocks) and gathered together in each block. 
In this process the transportations will be carried out 
by using the free motions 
of the direction perpendicular to the standing waves. 
It should be noticed that the vortex patterns which are a kind of 
topological property can be very stable in such transportations. 
Then they proceed to the next step carried out in the higher blocks, 
where interactions between vortices take place. 
Some vortices will be created and/or annihilated~\cite{k2}, 
and a new vortex pattern appears in each block. 
Such new patterns produced in the higher blocks will be sent 
to more higher blocks, 
where the higher step is performed. 
The same processes will be repeated, until the brain gets 
a set of vortex patterns that is recognized to be the final answer.  
Such hierarchy may be the base of our thinking processes. 
As we know, flows of fluids are very much flexible and easily 
deformed, but not easy to control them perfectly. 
Experiences and trainings 
in daily life must play a very important role 
in the construction of this hierarchy.  
At present we have so many unkown problems such as 
the definite correspondence between the vortex patterns and the recognitons, 
the dynamics in the higher blocks, and so on. 
We are still at the entrance of the understanding. 
The view given here, however, is possibly an idea for 
the understanding of thinking in terms of real physical processes.

\end {document}